\documentclass[conference]{IEEEtran}
\IEEEoverridecommandlockouts

\usepackage{cite}
\usepackage{amsmath,amssymb,amsfonts}
\usepackage{algorithm}
\usepackage{algpseudocode}
\usepackage{CJKutf8}
\usepackage{comment}
\usepackage{subcaption}
\usepackage{mathrsfs}
\usepackage{graphicx}
\usepackage{textcomp}
\usepackage{array}
\usepackage{xcolor}
\usepackage{soul, color}
\usepackage{xspace}
\usepackage{booktabs}
\usepackage{multirow}
\usepackage{graphicx}
\newcommand*{\note}[1]{\textcolor{red}{#1}}
\newcommand{\sysname}{$\textit{SBFL-LEO}$\xspace}
\def\BibTeX{{\rm B\kern-.05em{\sc i\kern-.025em b}\kern-.08em
    T\kern-.1667em\lower.7ex\hbox{E}\kern-.125emX}}
\begin{document}

\title{A Sharded Blockchain-Based Secure Federated Learning Framework for LEO Satellite Networks\\
}

\author{\IEEEauthorblockN{Wenbo Wu$^{1}$, Cheng Tan$^{1}$, Kangcheng Yang$^{1}$, Zhishu Shen$^{1}$\textbf{\IEEEauthorrefmark{1}}\thanks{\IEEEauthorrefmark{1} Corresponding author.}, Qiushi Zheng$^{2}$, Jiong Jin$^{2}$}


\IEEEauthorblockA{\textsuperscript{$^{1}$}School of Computer Science and Artificial Intelligence, Wuhan University of Technology, China\\
}
\IEEEauthorblockA{\textsuperscript{$^{2}$}School of Science, Computing and Engineering Technologies, Swinburne University of Technology, Australia\\
}

\IEEEauthorblockA{E-mail: \{wwb297985, cheng\_tan, yangkc\}@whut.edu.cn, z\_shen@ieee.org, \{qiushizheng, jiongjin\}@swin.edu.au}
}

\maketitle

\begin{abstract}

Low Earth Orbit (LEO) satellite networks are increasingly essential for space-based artificial intelligence (AI) applications. However, as commercial use expands, LEO satellite networks face heightened cyberattack risks, especially through satellite-to-satellite communication links, which are more vulnerable than ground-based connections. 
As the number of operational satellites continues to grow, addressing these security challenges becomes increasingly critical. Traditional approaches, which focus on sending models to ground stations for validation, often overlook the limited communication windows available to LEO satellites, leaving critical security risks unaddressed. To tackle these challenges, we propose a sharded blockchain-based federated learning framework for LEO networks, called \sysname. This framework improves the reliability of inter-satellite communications using blockchain technology and assigns specific roles to each satellite. Miner satellites leverage cosine similarity (CS) and Density-Based Spatial Clustering of Applications with Noise (DBSCAN) to identify malicious models and monitor each other to detect inaccurate aggregated models. Security analysis and experimental results demonstrate that our approach outperforms baseline methods in both model accuracy and energy efficiency, significantly enhancing system robustness against attacks.

\end{abstract}


\section{Introduction}

The advancement of satellite technology has driven the development of large Low Earth Orbit (LEO) satellite networks, with hundreds to thousands of satellites being launched. This trend has accelerated the commercialization of satellite-based Internet of Things (IoT) services and led to continuous upgrades in satellite technology \cite{ShenCSUR23}. As a result, modern satellites are now equipped with advanced cameras, processors, and antennas, enabling them to collect and process vast amounts of Earth imagery and sensor data through artificial intelligence (AI)-based solutions \cite{MahboobCST24}.
The traditional approach in LEO satellite networks relies on transmitting data to a central server. However, as data volumes grow, the centralized model training approach is becoming impractical due to high bandwidth costs, transmission delays, and the heightened vulnerability of satellite links compared to ground links~\cite{bao2021blockchain}.

Implementing blockchain~\cite{dong2024defending} and federated learning (FL)~\cite{mcmahan2017communication} offers an effective solution to this problem. In FL, each satellite aggregates locally calculated parameters and transmits model updates instead of raw data to jointly train a global model. Blockchain, as a decentralized, immutable, and traceable technology, eliminates the necessity of a central server in FL. Through its decentralized ledger, blockchain enables FL to transparently track updates and client operations across the entire network. For large-scale satellite networks, blockchain sharding technology is applied to enhance entire system performance~\cite{tang2023relayer,liu2023flexible}.  

Applying FL to LEO satellite networks still faces several challenges. One challenge is that FL assumes all distributed nodes are trustworthy, which is difficult to guarantee~\cite{chen2024credible}. Another challenge arises from the short, intermittent communication windows between satellites and ground data centers, making data transmission to the ground both time-consuming and often unnecessary\cite{razmi2022board}.

Several studies have been conducted to address the aforementioned challenges. For example, Wang \textit{et al.} proposed a method utilizing cosine similarity (CS) to filter malicious models by measuring the differences in model features\cite{wang2021edge}. However, as the FL model converges, the CS between the local model in the current round and the global model from the previous round increases significantly. This makes it essential to establish a CS threshold for accurate model classification. Chen \textit{et al.} divided the model into two categories by extracting model features and selected the model with the highest accuracy as the global model\cite{chen2024credible}.
This method can mistakenly classify some benign models as malicious in the absence of attacks, causing a reduction in accuracy. Zhu \textit{et al.} transmitted the models learned from satellites to the ground for model verification and aggregation~\cite{zhu2023privacy}. Although their methods effectively resist poisoning attacks, they did not consider the issue of short communication windows between satellites and the ground. With the increasing number of satellites, there is an urgent need for efficient and decentralized solutions that can operate directly within satellite networks~\cite{10399870}.


To address the aforementioned issues, we propose \sysname, a fully decentralized FL framework that integrates blockchain technology with sharding, enabling secure training on satellite networks. \sysname integrates CS and Density-Based Spatial Clustering of Applications with Noise (DBSCAN) to mitigate poisoning attacks. Specifically, model features are extracted by calculating the CS between each local model and the previous global model. DBSCAN then groups the models based on their CS values, automatically determining the number of clusters based on density thresholds and a minimum data point requirement. By enforcing a minimum cluster size, each cluster is limited to a maximum of two groups, enhancing resilience against poisoning attacks. The aggregated model with the highest accuracy is selected as the cluster model, effectively resisting poisoning attacks. In addition, satellites are categorized into three distinct roles, allowing the model trained by a learning satellite to be validated by a miner satellite rather than being transmitted to the ground.

The main contributions of this paper are summarized as follows:

\begin{itemize}

\item We propose \sysname, a blockchain-based federated learning framework to address the challenges of limited communication windows between satellites and ground stations while ensuring secure satellite communications. By assigning specific roles to satellites, \sysname avoids the need to transmit models to the ground for verification.


\item Within this framework, we introduce CS to extract model characteristics rather than explicitly rejecting models. DBSCAN is then applied to dynamically cluster models, preserving all benign models unaffected by compromise in the current round and reducing the accuracy loss caused by model rejection.


\item We evaluate \sysname using a real dataset and perform a thorough security analysis. Experimental results demonstrate that \sysname outperforms baseline methods in both learning accuracy and energy efficiency.

\end{itemize}

\section{System Model}

\subsection{Network Model}

\begin{figure}[tb!]
    \centering
    \includegraphics[width=0.95\linewidth]{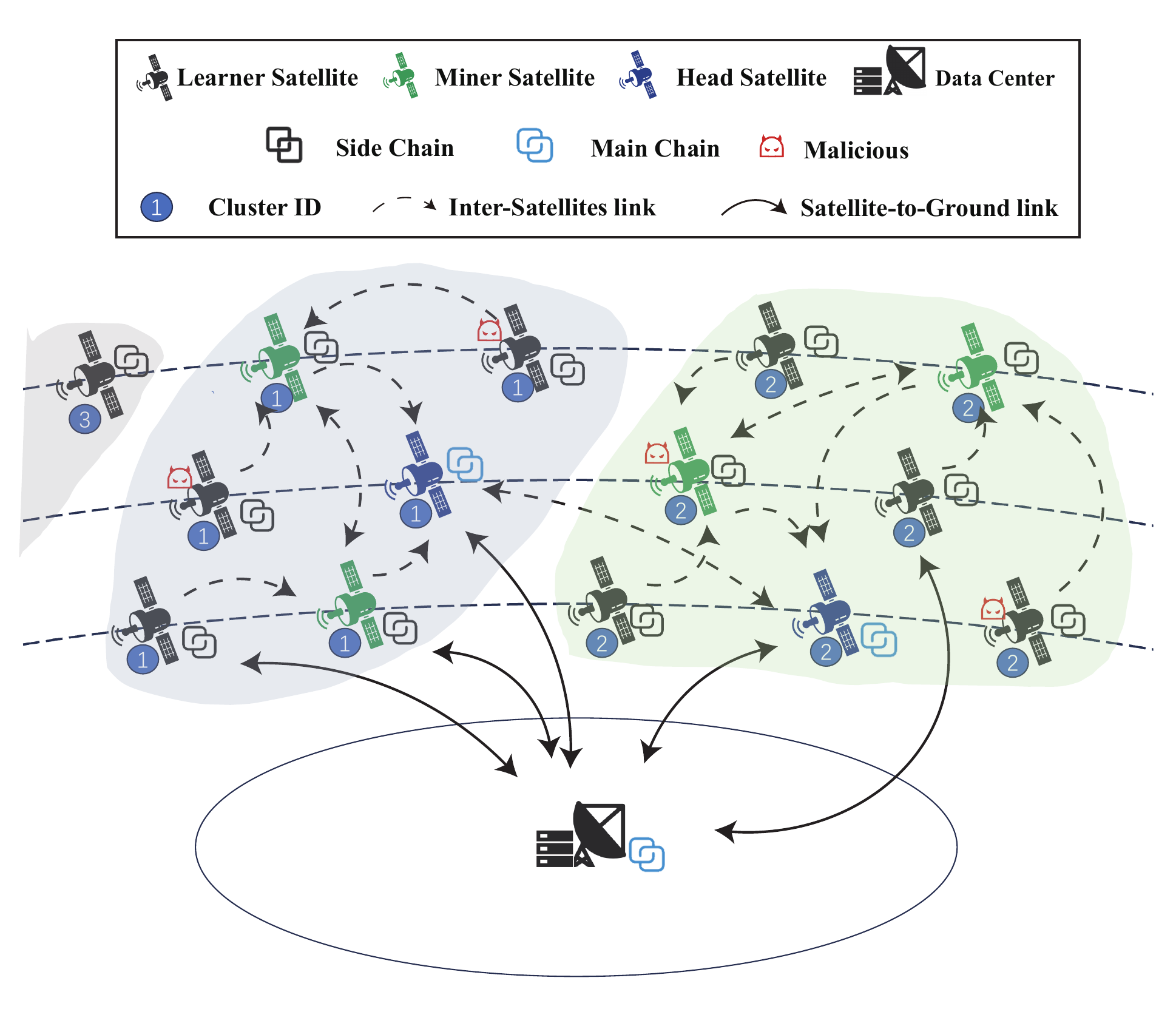}
    \caption{Network model.}
    \label{fig:model}
    \vspace{-0.4cm}
\end{figure}

We consider a LEO satellite system comprising both LEO satellites and a ground-based data center (DC).  As illustrated in \figurename~\ref{fig:model}, this system consists of a set of LEO satellites, each equipped with computational resources, denoted by $S=\{s_{1},s_{2},\cdots,s_{L}\}$ to support the processing of AI-based applications. Each satellite $s\in S$ possesses a local dataset $D_{s}$ of size $\lvert D_{s}\rvert$. Due to their fixed orbits, the data collected by satellites is often highly non-independent and evenly distributed. To mitigate the effects of data heterogeneity, DC separates satellites into numerous clusters $C^{r}=\{C^{r}_{1},C^{r}_{2}, \cdots,C^{r}_\mathcal{C}\}$ based on their position information and data characteristics. 

As shown in \figurename~\ref{fig:model}, we assume three functions for the satellites within each cluster: 
{\begin{itemize}
\item \textbf{Head satellite} is the  primary satellite within each cluster. It communicates with counterparts in other clusters to aggregate data and obtain the final global model for the current communication round. 
\item \textbf{Miner satellites} are the secondary satellites within the cluster, excluding the head satellite. Their responsibilities include evaluating each local model, performing classification aggregation, and voting on the aggregated model to select the representative local model for the cluster in a given communication round.
\item \textbf{Learner satellites}  update only their local models. 
\end{itemize}
}
Each satellite registers with the DC to obtain a blockchain account, and the data generated during the FL communication process is recorded on the blockchain, which is distributed across all devices. We used sharding technology to improve the performance of \sysname. Blockchain sharding significantly improves the scalability and processing efficiency of the system by dividing the blockchain network into multiple side chains. Each shard can independently process transactions and smart contracts, reducing network congestion and increasing transaction throughput. In addition, sharding can also share the burden of computation and storage, making blockchain more operable and efficient when facing large-scale applications. The satellites of each cluster are in the same side chain, and two main chains are maintained by the head satellite of each cluster.

Each head satellite maintains two main chains: the model chain, which stores the global model for each iteration along with each cluster's model, and the reputation chain, which records the reputation of the satellites. Within each cluster, miners manage a side chain that holds only the transactions generated within that cluster. Learning satellites maintain a single chain dedicated to storing their locally generated models. As lightweight nodes, learner satellites do not store the entire blockchain. Instead, they keep only the block headers and utilize Merkle proofs to verify the inclusion of transactions in the current blockchain transaction list.


\subsection{Communication Model}

We assume that the satellite system comprises two types of communication links: the Satellite-to-Ground Link (SGL) and the Inter-Satellite Link (ISL). The SGL is used to transmit clustering results from DC to the satellites at the beginning of the FL process, while the ISL facilitates the transfer of model parameters and lists of suspicious satellites among the satellites. Specifically, within cluster $C^{r}_{i}$, each learner satellite connects to the nearest miner satellite, which then communicates with all other miner satellites. The effective path loss between nodes $u$ and $v$ can be expressed as\cite{xiong2024energy}:
\begin{equation}
   L_{uv} = \left( \frac{c}{4 \pi d_{uv} f_c} \right)^2, \quad \forall u, v \in {C^{r}_{i} \cup DC},
\end{equation}
where $c$ is the speed of light, $f_c$ is the carrier frequency used for the link, and $d_{uv}$ is the physical distance between the two communicating nodes. The signal-to-noise ratio (SNR) for the ISL between satellites $u$ and $v$ can defined as:
\begin{equation}
    \gamma_{uv} = {G^{tr}_ {uv} G^{re}_{uv} L_{uv}}/{N_0},
\end{equation} 
where $G^{tr}_ {uv}$ is the gain of the transmitting antenna for satellite $u$ toward satellite $v$,  $G^{re}_{uv}$ is the gain of the receiving antenna for satellite $v$ from satellite $u$, $N_0$ is the noise power spectral density. The SNR of SGL between DC and satellite $v$ can be defined as:
\begin{equation}
    \gamma_{{\scriptscriptstyle DC}-v} = {G^{tr}_ {{\scriptscriptstyle DC}-v} G^{re}_{{\scriptscriptstyle DC}-v} L_{{\scriptscriptstyle DC}-v}}L_a/{N_0}.
\end{equation} 
Herein $L_a$ represents the additional loss introduced by the meteorological environment.

The achievable data transmission rate for SGL and ISL can be obtained by:
\begin{equation}
   R_{uv} = B \log_2 \left( 1 + \gamma_{uv} P_u \right), 
   \label{eq:rate}
\end{equation}
where $B$ is the link bandwidth and $P_{uv}$ is the power transmitted from node $u$ to node $v$.

\subsection{Energy Consumption Model}
The FL process can be divided into three stages: model distribution, federated learning within the cluster, and model aggregation between the head satellites across clusters. The energy consumption model is outlined as follows.

\subsubsection{Model distribution}
At the beginning of each communication round, the head satellite distributes the aggregated global model from the previous round to the other satellites within the cluster. Specifically, the head satellite first transmits the model to the miner satellites, which then relay it to the associated learner satellites. In cluster $C_i$, the energy consumption for distributing the model to the other satellites can be calculated as follows:
\begin{equation}
   E_{di}^{c_i} = \sum_{p_{uv} \in P_{C_i}} \frac{P_u |w_g|}{R_{uv}}, 
   \label{eq:v}
\end{equation}
where $\lvert w_g \rvert$ is the size of the global model parameters.

\subsubsection{Federated learning within the cluster}
When learner $l$ receives the global model transmitted by the miner, it begins training its local model using its dataset.  Let $\phi$ represent the number of CPU cycles required to process a single data sample. The time required for learner $l$ to complete the training can be calculated as follows:
\begin{equation}
   T_{cmp}^l = \frac{\tau \phi |D_l|}{f_l}, 
\end{equation}
where $f_l$ is the computing frequency allocated to learner $l$ for local training. Then the energy consumption required for model training can be calculated using the following formula:
\begin{equation}
   E_{cmp}^l = \epsilon_0 f_l^3 T_{cmp}^l
\end{equation}
where $\epsilon_0$ is the constant coefficient determined by the hardware architecture. Learner $l$ then transmits its local model $w_l$ to the associated miner, which verifies the signature before forwarding it to the other miners. The energy consumption for transmission $E_{tr}^{c_i}$ can be calculated using Formula (\ref{eq:v}).

Subsequently, the miners group the local models with similar data distributions using clustering techniques such as CS and DBSCAN.
Next, miner $m$ trains the models for $\tau$/2 epochs using local dataset $D_m$ and selects the model with the highest accuracy to send to the head. The energy consumption for this process can be calculated as follows:
\begin{equation}
    E_{ev}^m = \frac{\tau \epsilon_0 f_m^2 \phi |D_m|}{2}.
    \label{eq:evl}
\end{equation}

The head $h_i$ selects the model that receives the highest number of votes as the aggregated model $w^i$ for the cluster.

The energy consumption $E_{c_i}$ within cluster $c_i$ during a communication round can be calculated using the following formula:
\begin{equation}
    E_{c_i} = \sum_{l \in L_i}E_{cmp}^l  + \sum_{m \in M_i}E_{ev}^m + E_{tr}^{c_i}.
\end{equation}

\subsubsection{Model aggregation across clusters}
After the aggregation within the cluster is complete, $h_i$ transmits the model $w^i$ to the other head satellites. The energy consumption required for inter-cluster transmission $E_{tc}^r$ can be calculated as follows:
\begin{equation}
    E_{tc}^r = \sum_{p_{h_1h_2} \in P_{H^r}}\frac{P_{h_1} |w_{h_1}|}{R_{h_1h_2}}.
\end{equation}

Upon receiving models from other clusters, the head satellite verifies the accuracy of each model using its local dataset. The energy consumption for the verification process $E_{ve}^r$ can be calculated using Formula (\ref{eq:evl}). If the accuracy error is below a predefined threshold value $\sigma$, the head satellite votes in favor of the model; otherwise, the model is rejected.
The energy consumption for global round $r$ is:
\begin{equation}
    E^r = \sum_{c_i \in C^r} \left( E_{di}^{c_i} + E_{c_i}\right) + E_{tc}^r + \mathcal{C}E_{ve}^r,
\end{equation}

\section{Overview of \sysname}

\subsection{Working mechanism}
\begin{figure}[tb!]
    \centering
    \includegraphics[width=0.95\linewidth]{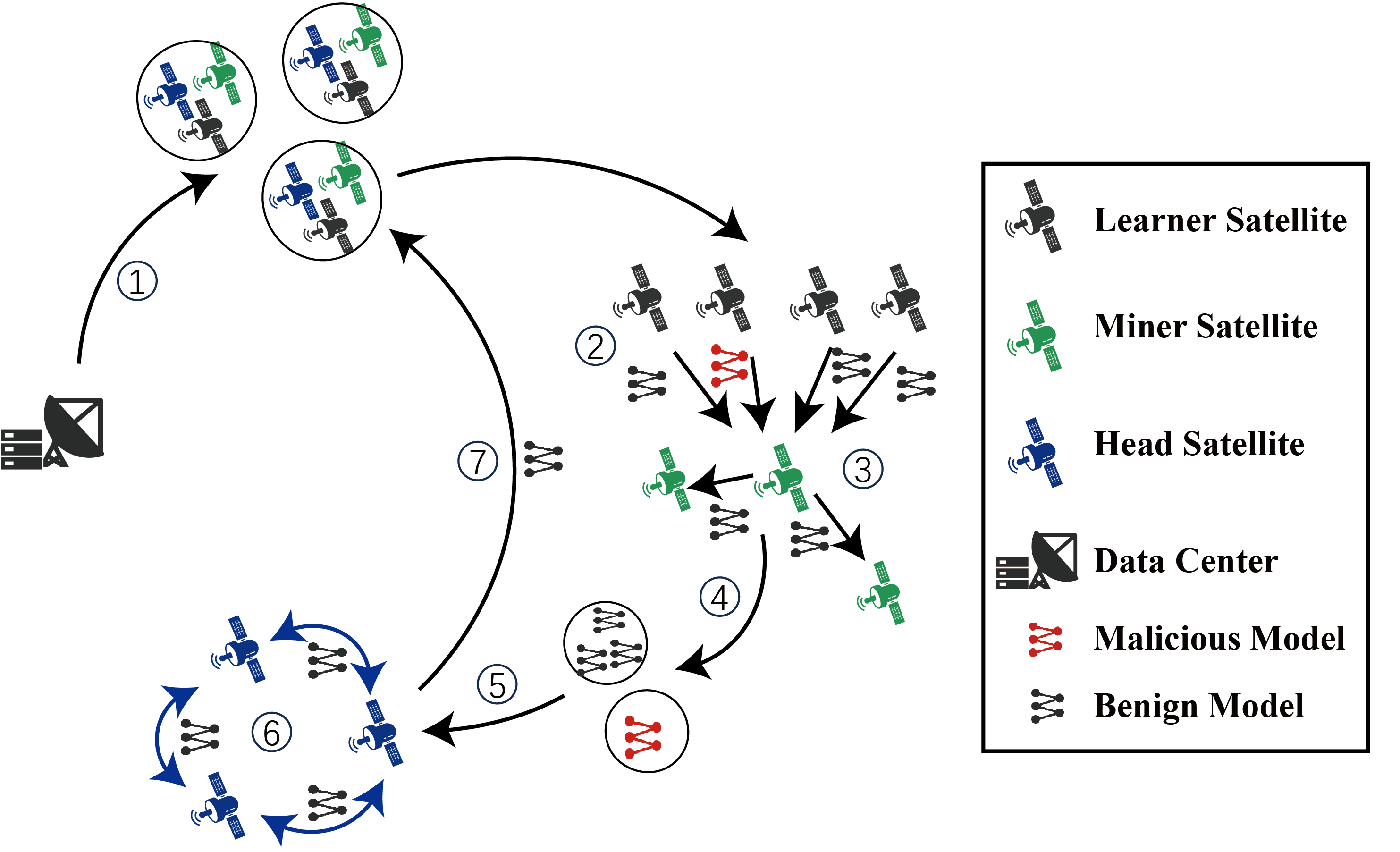}
    \caption{Working mechanism of \sysname.}
    \label{fig:workflow}
\end{figure}
We propose \sysname, a sharded blockchain-based satellite federated learning framework to mitigate malicious attacks throughout the FL process. The workflow of \sysname is shown in \figurename~\ref{fig:workflow}. The DC groups satellites based on geographic location and data characteristics and then assigns roles according to each satellite’s reputation. Learner satellites train models and transmit them to miner satellites (Step 2 to 3 in \figurename~\ref{fig:workflow}). After transmitting the local model to other miner satellites 
, each miner satellite classifies the local model and selects the model with the highest accuracy to transmit to the head satellite (Step 4 to 5 in \figurename~\ref{fig:workflow}). Once the global model for the communication round is obtained through inter-cluster consensus among head satellites, it is distributed to other satellites within the cluster. The process from Step 2 to Step 7 in \figurename~\ref{fig:workflow} will repeat until the global model achieves the desired level of accuracy.

\sysname consists of two key components: intra-cluster consensus and inter-cluster consensus. The details are outlined below:

\subsection{Intra-cluster Consensus}
The satellites within the $C^{r}_{i}$ cluster are categorized into a learner set $L_{i}=\{l_{1},l_{2},\cdots,l_{L}\}$ and a miner committee $M_{i}=\{m_{1},m_{2},\cdots,m_{M}\}$, where $ C^{r}_{i} $ = $L_{i} \cup M_{i}$. The set of head satellites can then be denoted as $H^{r} = \{h^{r}_{1},h^{r}_{2},\cdots, h^{r}_\mathcal{C}\}$. The objective of each learner satellite $l\in L_{i}$ is to minimize the global loss function by using its local data for $\tau$ epochs to train the global model $w^{r-1}$ from the previous communication round. We have added a penalty term based on satellite energy consumption for model training in the loss function, which can enable satellites with lower energy consumption to make greater contributions. The loss function $F_l(w)$ for the satellite $l$ is given by:
\begin{equation}
   F_l(w^r_l) = \frac{1}{|D_l|} \sum_{\xi \in D_l} f(w^r_{l}, \xi) + \lambda E_{\text{cmp}}^l,
\end{equation}
where $f$($w^r_{l}$, $\xi$) represents the loss function of the learner satellite's local model at data point $\xi$, and $\lvert D_{l} \rvert$ is the size of the dataset for learner satellite $l$. For $\tau$ epochs, each satellite trains its model locally to reduce communication costs. The local model is then updated by the learner satellite $l$ using the stochastic gradient descent (SGD) method with a learning rate $\eta$ as follows:
\begin{equation}
   w_l^r = w_l^{r-1} - \eta \nabla F_l(w^{r-1}_l).
\end{equation}

After $\tau$ epochs, each learner satellite submits its trained local model to the miner committee for verification and aggregation to generate the cluster model. When the learning satellite $l$ completes its local training, it generates a transaction $t^r_l$ containing its model $w^r_l$, the size of its dataset $|D_l|$ and signs it with its private key. This satellite then transmits the signed transaction to the nearest miner satellite $m$. Upon receiving the transaction $t^r_l$, miner satellite $m$ calculates the CS between the local model $w^r_l$ and the global model $w^{r-1}$ from the previous round. The formula for calculating the CS $\theta\left(w^r_l\right)$ is as follows:
\begin{equation}
    \theta\left(w^r_l\right) = \frac{w^r_l \cdot w^{r-1}}{\|w^r_l\| \cdot \|w^{r-1}\|}.
\end{equation}

The miner satellite $m$ then uses DBSCAN to group these models:
\begin{equation}
    G[1, \ldots, \mathcal{G}] = \textit{DBSCAN}\left(\theta(w^r_1), \theta(w^r_2),\dots, \theta(w^r_L)\right).
\end{equation}

Since satellites in close proximity often capture similar data, the model parameters trained on their local datasets tend to have a high CS with the global model. DBSCAN can group these models together while filtering out poisoned models.

The miner satellite calculates the weight of each learner satellite's local model based on the size of each learner satellite dataset and the communication rate with the learning satellite. Each miner satellite then trains the aggregated model for $\tau/2$ epochs using its training dataset to determine the accuracy score for each group of models. The miner satellite selects the model $w^r_m$ with the highest accuracy. Model aggregation is achieved using the following formula:
\begin{equation}
    w_{G_g} = \frac{\sum R_{uv} \cdot |D_l| \cdot w_l^r}{\sum R_{uv} \cdot |D_l|}, \quad \forall w_l^r \in G_g.
\end{equation}

Learner satellites that do not belong to the group of the selected model are added to the suspect list $\mathcal{L}^r_m$. After miner satellite $m$ selects the model, it generates a transaction $t^r_m$ that includes its voting result, the accuracy score $e^r_{m \rightarrow w^r_m}$, and the list $\mathcal{L}^r_m$, which it then sends to the head satellite $h_i$.

Upon receiving transactions from all miner satellites within the cluster, the head satellite tallies the votes and selects the satellite with the highest number of votes as the cluster model $w^r_i$. Miner satellites that submit incorrect votes are also added to the head satellite's suspect list $\mathcal{L}^r_{h_i}$. At this point, the model update within the cluster is complete.

\subsection{Inter-cluster Consensus}
After completing the intra-cluster update, the head satellite $h_i$ generates a transaction $t^r_{h_i}$ containing the cluster model parameters $w^r_{h_i}$, the accuracy score of the model $e^r_{{h_i} \rightarrow w^r_{h_i}}$, and the list of suspicious nodes $\mathcal{L}^r_{h_i}$. This transaction is signed by $h_i$ and sent to the head satellites of other clusters. Upon receiving transactions from other head satellites, head satellite $h_i$ trains each cluster model on its local dataset for $\tau/2$ epochs to evaluate the accuracy score for each model $e^r_{h_i} = \{ e^r_{{h_i} \to w^r_{h}}, \, h \in H^r \}$. If the difference between $e^r_{{h_i} \to w^r_{h}}$ and $e^r_{h \to w^r_{h}}$ is less than $\sigma$, $h$ approves the model $w^r_{h}$ for participation in the global model aggregation; otherwise, it is rejected. Next, head satellite $h_i$ encapsulates the voting results and the model's accuracy score $e^r_{h_i}$ into transaction $t_{h_i}^{vot}$ and sends it to other head satellites.

After summing the voting results from all head satellites, the accepted models are aggregated, based on their accuracy scores, to get the global model $w^r_g$ for communication around $r$. Head satellites that fail to cast their votes correctly, along with the satellites in their cluster that trained on the corresponding local models, are also added to the suspicious list $\mathcal{L}^r$. Additionally, the suspicious list $\mathcal{L}^r$ includes the list of suspicious satellites from the honest head satellites. Finally, the head satellite uses an aggregation method based on accuracy weights to aggregate the global model, as described by the following formula:
\begin{equation}
   w^r_g = \frac{\sum e^r_{hi \to w^r_{h}} w_h^r}{\sum e^r_{hi \to w^r_{h}}},\quad \forall (h,h_i \in H_r) \land (h,h_i \notin \mathcal{L}^r).
   \label{eq:agg}
\end{equation}

The loss function of the global model is determined by:
\begin{equation}
   Loss_{global}(w^r_g) = \frac{1}{\mathcal{C}} \sum_{i=1}^{\mathcal{C}} w_i L_i.
\end{equation}

After completing a round of FL, each learner satellite that updates a benign model and each miner satellite that casts a correct vote receives a reputation reward. Head satellites that act honestly also receive rewards, allowing them to retain their roles in the next round. This approach aims to keep head satellites consistent, reducing the need for large-scale blockchain data transfers to newly appointed head satellites. Satellites in the list $\mathcal{L}^r$ are penalized with a reputation reduction, diminishing their impacts on FL in subsequent communication rounds.

\sysname aims to minimize the loss function to optimize $w$. Throughout the entire FL process, satellites identified as suspicious nodes are penalized through reputation deduction, which continues until they are eventually removed from the system. FL proceeds until a global model that minimizes the loss function is obtained, as follows:
\begin{equation}
    w = \arg \min_{w^r_g} L_{\text{global}}(w^r_g).
\end{equation}


\section{Performance Analysis}

\subsection{Experiment Settings}
We consider a LEO satellite network with 200 satellites, evenly distributed across 20 orbits. Moreover, we set the maximum communication power as 5~W, the maximum computing frequency as 5~GHz, the link bandwidth as 20 MHz, $\gamma_{uv}$ = $10^3 $,  $\epsilon_0$ = $10^{-28}$, and $\phi$ = $10^5$ cycles/sample\cite{zhai2023fedleo}. We first implement FL using the FedAvg~\cite{mcmahan2017communication} by adopting a convolutional neural network (CNN) with a learning rate $\eta=0.1$ for local training, leveraging CNN's effectiveness in image classification tasks. The model is tested on the MNIST dataset with a training/testing ratio of 80:20. In this FL framework, we set the number of local training epochs to 20, the batch size to 64, the proportion of malicious clients in the network to $20\%$, and the number of cluster $\mathcal{C}$ is 5.

We validate the effectiveness of our proposed \sysname against the benchmark Weighted FedAvg. We introduce two methods based on FedAvg: \textbf{FedAvg\_with\_M} with 20\% malicious satellites and \textbf{FedAvg} without any malicious attacks. In addition, we develop \textbf{SBFL-LEO-k\_means} that uses $k$-means algorithm to classify the CS of the model for ablation study, and \textbf{eFL} that directly filtering malicious models through CS~\cite{wang2021edge}.

\begin{figure*}[t]
	\begin{minipage}[b]{.63\columnwidth}
		\centering
	\includegraphics[width=\columnwidth]{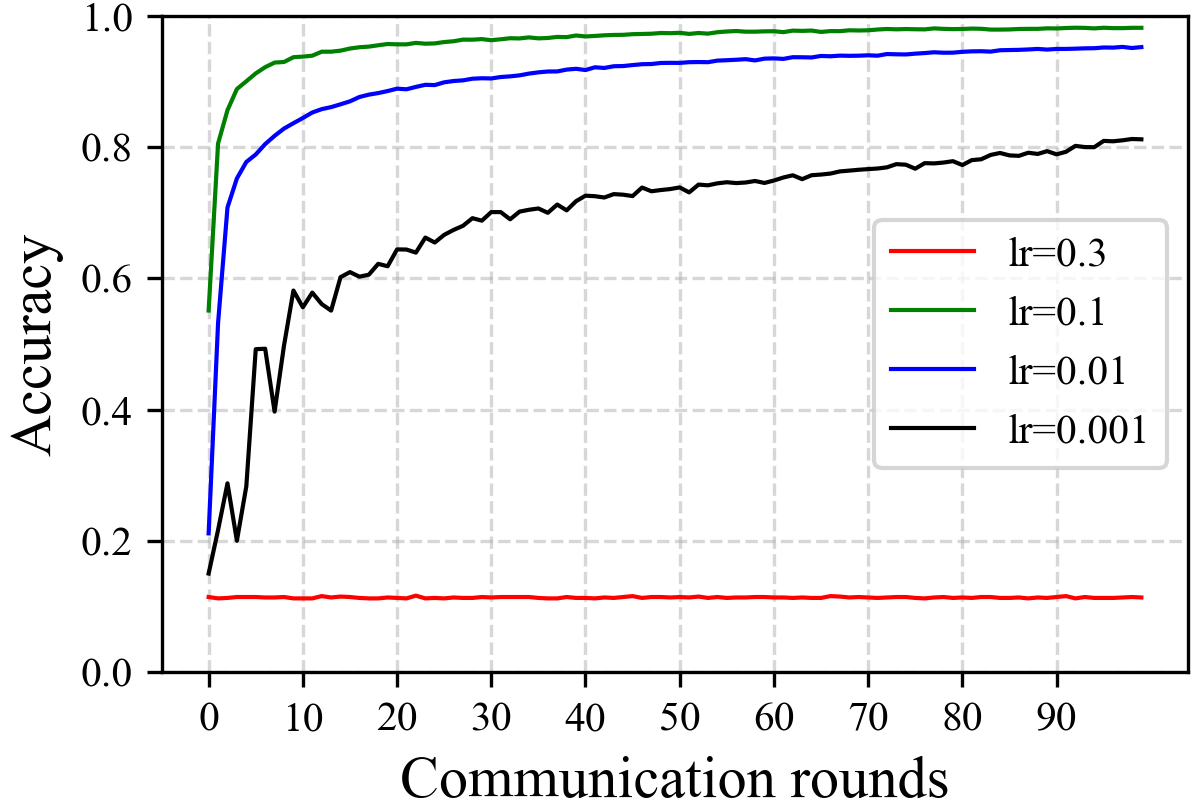}
		\subcaption{Accuracy with different learning rates
  }\label{fig:lr}
	\end{minipage}
 	\centering
	\begin{minipage}[b]{.63\columnwidth}
		\centering
	\includegraphics[width=\columnwidth]{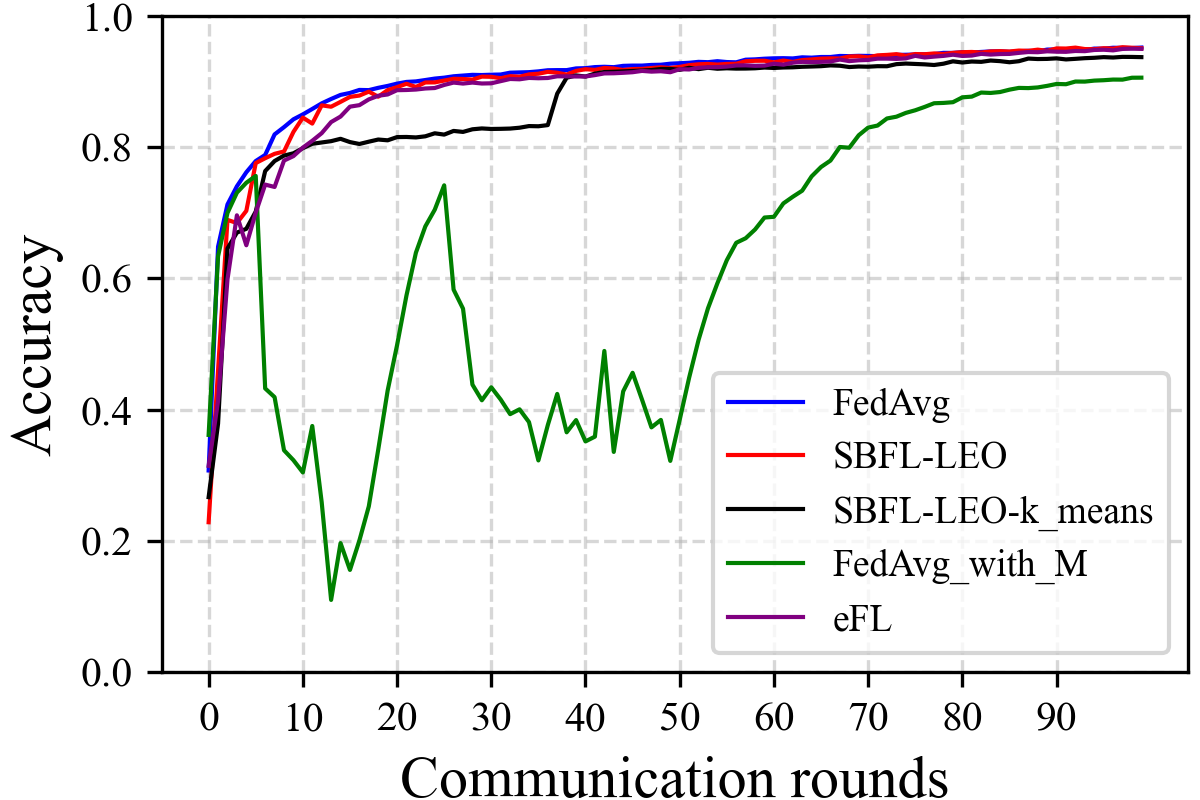}
		\subcaption{Accuracy for different methods}\label{fig:dif_method}
	\end{minipage}
 	\centering
	\begin{minipage}[b]{.63\columnwidth}
		\centering
	\includegraphics[width=\columnwidth]{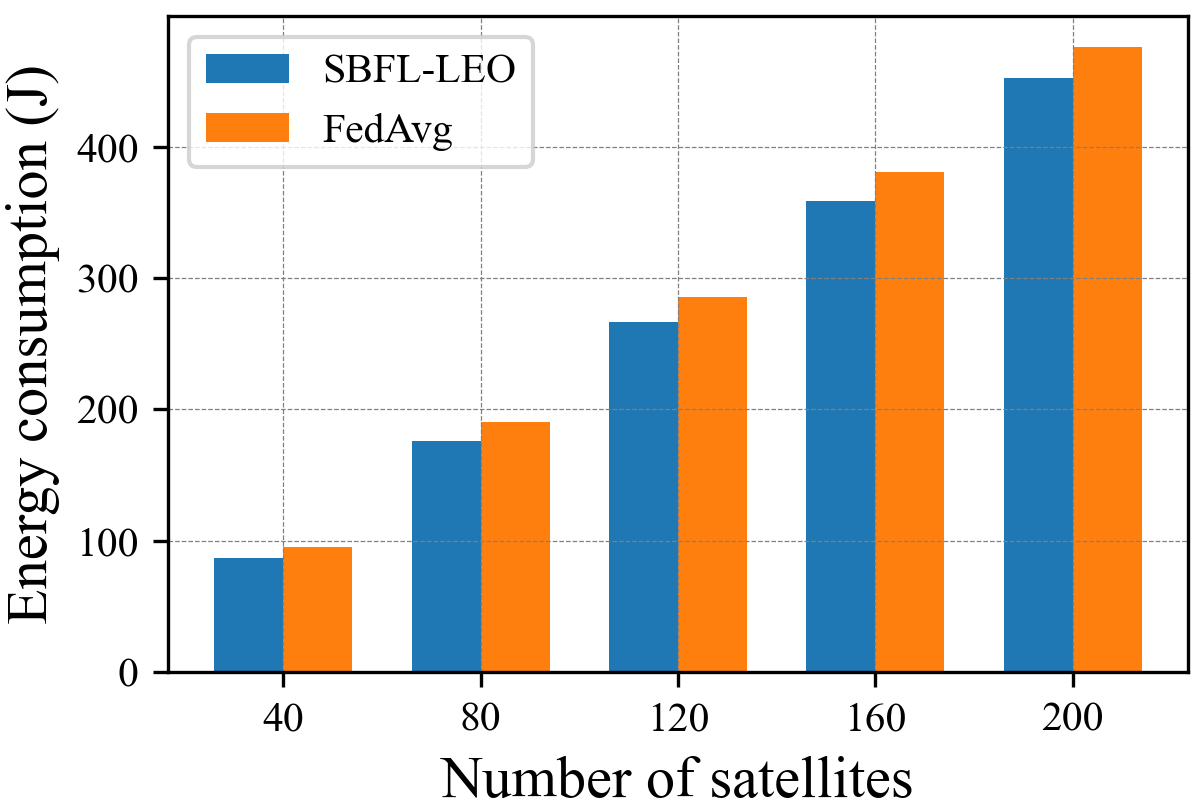}
		\subcaption{Energy consumption}\label{fig:energy}
	\end{minipage}
        \caption{Model accuracy and energy consumption performance.}
	\label{fig:result1}
 \vspace{-0.2cm}
\end{figure*}

\subsection{Experimental Results Analysis}

We test the model at different learning rates, and the results are shown in \figurename~\ref{fig:lr}. When the learning rate (\textit{lr}) is set to 0.1, the model converges faster and achieves superior performance. In contrast, a learning rate of 0.3 results in the model failing to update due to the high rate. At a learning rate of 0.01, the model's accuracy is 3\% lower than when the learning rate is 0.1. Meanwhile, at a learning rate of 0.001, the slow model converge speed makes it challenging to reach optimal accuracy.

\figurename~\ref{fig:dif_method} illustrates the results of model accuracy across various methods. We assume malicious clients upload poisoned local models in rounds 5$\sim$15 and 25$\sim$50 to simulate targeted attacks within the FL process. Due to the inability to identify malicious models, the accuracy of the FedAvg\_with\_M decreases by over 50\% in 5$\sim$15 rounds and there is no improvement in accuracy between rounds 25$\sim$50. FedAvg, eFL, SBFL-LEO-k\_means and \sysname's accuracy continue to increase with the increase of communication rounds. EFL, our proposed \sysname and SBFL-LEO-k\_means lose some local models by designating certain satellites as miner satellites, resulting in lower accuracy compared to FedAvg. Furthermore, the lower accuracy of SBFL-LEO-k\_means compared to our method stems from its inability to dynamically adjust the number of model groups, which leads to the exclusion of some benign models during rounds without malicious attacks. During attacks in rounds 5-15, eFL exhibits an accuracy difference of approximately 4\% compared to \sysname. This is because eFL does not consider the scenarios where malicious satellites submit insufficiently trained models or even directly submitted the previous round's global model while utilizing CS to filter out harmful models, ultimately hindering the convergence speed of the global model. In contrast, DBSCAN enables dynamic regrouping, which helps prevent the loss of benign models. In the 20th round, the accuracy difference between \sysname and FedAvg is less than 1\%. Therefore, our proposed \sysname effectively defends against poisoning attacks and preserves benign models during rounds without malicious attacks. 

As illustrated in \figurename~\ref{fig:energy}, the energy consumption per communication round grows with the number of satellites. However, our proposed \sysname consumes less energy compared to FedAvg. For instance, when the number of satellites reaches 120, the energy consumption for our proposed strategy is 266.34~$J$, while FedAvg consumes 285.48~$J$. This reduction in energy usage can be ascribed to our proposal \sysname, where 20\% of the satellites are designated as miner satellites. These miner satellites do not undertake local training but instead validate the models of the learning satellites, hence decreasing overall energy consumption. 

Table \ref{tab:time} analyzes the time consumption for different methods when achieving 90\% model accuracy. When continuously subjected to malicious attacks, FedAvg\_with\_M fails to converge due to its inability to resist poisoning attacks. eFL converges slower than \sysname because it cannot exclude models with insufficient training. As a result of losing some effective models during the training process, the aggregation speed of SBFL-LEO-k\_means is slower than that of \sysname. Therefore, our proposed \sysname effectively defends against poisoning attacks while generating high-performance global models. Furthermore, our method achieves improvements in energy efficiency.


\begin{table}[t]
\scriptsize{
\centering
\caption{Comparisons of time cost for different methods.}
\label{tab:time}
\begin{tabular}{>{\centering\arraybackslash}m{2.5cm} cc}
\toprule
\multirow{2}{*}{\textbf{Methods}} & \multicolumn{2}{c}{\textbf{Average time to achieve accuracy of 90\%, and}} \\
& \multicolumn{2}{c}{\textbf{ the number of communication rounds to achieve}} \\
\cmidrule(lr){2-3}
& Average time (s)  & Communication rounds \\
\midrule
\sysname & 6.415 & 22$\pm$2 \\
eFL & 9.331 & 32$\pm$2 \\
SBFL-LEO-k\_means & 11.664 & 40$\pm$2 \\
FedAvg\_with\_M & $\infty$ & $\infty$ \\
\bottomrule
\end{tabular}
}
\vspace{-0.3cm}
\end{table}

\subsection{Security Analysis}
In this subsection, we discuss whether the miner committee can achieve consensus to obtain the correct aggregated model. After classifying the local models, miners validate and evaluate the aggregated models of each group, and then vote for the model with the highest accuracy. However, malicious miners may vote dishonestly, favoring models with lower accuracy.  Since the model with the majority vote is ultimately selected, the cluster can still accurately complete the aggregation as long as malicious miners constitute less than 50\% of the group. This conclusion also applies to inter-cluster aggregation, enabling the identification of clusters that have failed to select a benign model. Therefore, it is essential that over 50\% of the head satellites are benign at the beginning of the first round to guarantee the reliable progression of federated learning.

\section{Conclusion}
In this paper, we propose \sysname, a federated learning framework based on sharded blockchain, designed to defend against malicious attacks in LEO satellite networks while considering satellite energy consumption. Specifically, we utilize blockchain technology to ensure secure communication between satellites. Considering the issue of short communication windows between satellites and the ground, we categorize satellites into roles, and mine satellites use CS and DBSCAN based methods to detect malicious learning satellites, and supervise miners through inter-cluster consensus and intra-cluster consensus. Experimental results and security analysis demonstrate that \sysname effectively mitigates malicious attacks, reduces system energy consumption, and maintains satisfactory model accuracy performance.

\bibliographystyle{ieeetr}
\bibliography{ref.bib}

\begin{thebibliography}{10}

\bibitem{ShenCSUR23}
Z.~Shen {\em et~al.}, ``A survey of next-generation computing technologies in space-air-ground integrated networks,'' {\em ACM Computing Surveys}, vol.~56, no.~1, 2023.

\bibitem{MahboobCST24}
S.~Mahboob and L.~Liu, ``Revolutionizing future connectivity: A contemporary survey on {AI}-empowered satellite-based non-terrestrial networks in {6G},'' {\em IEEE Communications Surveys \& Tutorials}, vol.~26, no.~2, pp.~1279--1321, 2024.

\bibitem{bao2021blockchain}
Z.~Bao, M.~Luo, H.~Wang, K.-K.~R. Choo, and D.~He, ``Blockchain-based secure communication for space information networks,'' {\em IEEE Network}, vol.~35, no.~4, pp.~50--57, 2021.

\bibitem{dong2024defending}
N.~Dong, Z.~Wang, J.~Sun, M.~Kampffmeyer, W.~Knottenbelt, and E.~Xing, ``Defending against poisoning attacks in federated learning with blockchain,'' {\em IEEE Transactions on Artificial Intelligence}, vol.~5, no.~7, pp.~3743--3756, 2024.

\bibitem{mcmahan2017communication}
B.~McMahan, E.~Moore, D.~Ramage, S.~Hampson, and B.~A. y~Arcas, ``Communication-efficient learning of deep networks from decentralized data,'' in {\em Proceedings of the International Conference on Artificial Intelligence and Statistics (AISTATS)}, pp.~1273--1282, 2017.

\bibitem{tang2023relayer}
H.~Tang {\em et~al.}, ``Relayer-enabled sharding blockchain for satellite internet with high concurrency,'' in {\em Proceddings of the IEEE Global Communications Conference (GLOBECOM)}, pp.~7580--7585, 2023.

\bibitem{liu2023flexible}
Y.~Liu {\em et~al.}, ``A flexible sharding blockchain protocol based on cross-shard {Byzantine} fault tolerance,'' {\em IEEE Transactions on Information Forensics and Security}, vol.~18, pp.~2276--2291, 2023.

\bibitem{chen2024credible}
L.~Chen {\em et~al.}, ``A credible and fair federated learning framework based on blockchain,'' {\em IEEE Transactions on Artificial Intelligence}, pp.~1--15, 2024.

\bibitem{razmi2022board}
N.~Razmi, B.~Matthiesen, A.~Dekorsy, and P.~Popovski, ``On-board federated learning for dense {LEO} constellations,'' in {\em Proceedings of the IEEE International Conference on Communications (ICC)}, pp.~4715--4720, 2022.

\bibitem{wang2021edge}
T.~Wang {\em et~al.}, ``Edge-based communication optimization for distributed federated learning,'' {\em IEEE Transactions on Network Science and Engineering}, vol.~9, no.~4, pp.~2015--2024, 2021.

\bibitem{zhu2023privacy}
J.~Zhu, J.~Wu, A.~K. Bashir, Q.~Pan, and W.~Yang, ``Privacy-preserving federated learning of remote sensing image classification with dishonest majority,'' {\em IEEE Journal of Selected Topics in Applied Earth Observations and Remote Sensing}, vol.~16, pp.~4685--4698, 2023.

\bibitem{10399870}
X.~Luo, H.-H. Chen, and Q.~Guo, ``{LEO/VLEO} satellite communications in {6G} and beyond networks–technologies, applications, and challenges,'' {\em IEEE Network}, vol.~38, no.~5, pp.~273--285, 2024.

\bibitem{xiong2024energy}
T.~Xiong {\em et~al.}, ``Energy-efficient federated learning for earth observation in {LEO} satellite systems,'' in {\em Proceedings of the IEEE Wireless Communications and Networking Conference (WCNC)}, pp.~01--06, 2024.

\bibitem{zhai2023fedleo}
Z.~Zhai, Q.~Wu, S.~Yu, R.~Li, F.~Zhang, and X.~Chen, ``{FedLEO}: An offloading-assisted decentralized federated learning framework for low earth orbit satellite networks,'' {\em IEEE Transactions on Mobile Computing}, vol.~23, no.~5, pp.~5260--5279, 2023.

\end{thebibliography}

\end{document}